\documentclass[final,3p,times,twocolumn,authoryear]{elsarticle}

\usepackage{amssymb}
\usepackage{xcolor}

\usepackage{lineno}

\usepackage{rotating}

\journal{Icarus}

\begin{document}

\begin{frontmatter}



\title{A Collisional Origin to Earth's Non-chondritic Composition?}

\author[label1]{Amy Bonsor}
\author[label1]{Zo\"e M. Leinhardt}
\author[label1]{Philip J. Carter}
\author[label2]{Tim Elliott}
\author[label2]{Michael J. Walter}
\author[label3]{Sarah T. Stewart}

\address[label1]{School of Physics, H.H. Wills Physics Laboratory, University of Bristol, Tyndall Avenue, Bristol BS8 1TL, UK}
\address[label2]{School of Earth Sciences, University of Bristol, Bristol, BS8 1RJ, UK}
\address[label3]{Department of Earth and Planetary Sciences, University of California, Davis, One Shields Avenue, Davis, California 95616, USA}

\begin{abstract}
Several lines of evidence \textcolor{black}{indicate} a non-chondritic composition \textcolor{black}{for} Bulk Earth. If Earth formed from the accretion of chondritic material, its non-chondritic composition, in particular the super-chondritic $^{142}\mathrm{Nd}/^{144}\mathrm{Nd}$ and low Mg/Fe ratios, \textcolor{black}{might} be explained by the collisional erosion of differentiated planetesimals during its formation. In this work we use \textcolor{black}{an} $N$-body code, that includes a state-of-the-art collision model, to follow the formation of protoplanets, similar to proto-Earth, from differentiated planetesimals ($>$ 100 km) up to isolation mass ($>$ 0.16 M$_\oplus$). Collisions between differentiated bodies have the potential to change the core-mantle \textcolor{black}{ratio} of the \textcolor{black}{accreted} protoplanets. We show that sufficient mantle material can be stripped from the colliding bodies during runaway and oligarchic growth, such that the final protoplanets could have Mg/Fe and Si/Fe ratios similar to that of bulk Earth, but only if Earth is an extreme case and the core is assumed to contain 10\% silicon by mass. This may indicate an important role for collisional differentiation during the giant impact phase if Earth formed from chondritic material.
\end{abstract}


\end{frontmatter}


\section{Introduction}
\label{}

\textcolor{black}{A basic premise when estimating elemental budgets of planets is that they were constructed from material represented by our collection of undifferentiated, chondritic meteorites. This assumption has recently come under close scrutiny due to the super-chondritic $^{142}\mathrm{Nd}/^{144}\mathrm{Nd}$ composition of the accessible Earth \citep{Boyet:2005}. This observation has generally been explained by one of two scenarios for generating elevated $^{142}\mathrm{Nd}/^{144}\mathrm{Nd}$ as a result of radiogenic ingrowth from the short-lived $^{146}\mathrm{Sm}$}.\footnote{\textcolor{black}{For example, $^{142}\mathrm{Nd}/^{144}\mathrm{Nd} = (^{142}\mathrm{Nd}/^{144}\mathrm{Nd})^\circ + (^{144}\mathrm{Sm}/^{144}\mathrm{Nd})(^{146}\mathrm{Sm}/^{144}\mathrm{Sm})^\circ(1-e^{-\lambda t_1})$ for a reservoir formed as a closed system $t_1$ years after the start of the solar system with super chondritic $^{144}\mathrm{Sm}/^{144}\mathrm{Nd}$ (proportional to bulk Sm/Nd). The symbol $^\circ$ indicates initial values at the start of the solar system inferred from primitive meteorites, $\lambda$ is the decay constant of $^{146}\mathrm{Sm} \sim 1.02\times10^{-7} yr^{-1}$ \citep{Kinoshita:2012}.}}
\textcolor{black}{Namely, either the Earth has an untapped, hidden reservoir, most likely located at the bottom of the mantle \citep{Boyet:2005, Labrosse:2007}, leaving the observable, outer silicate Earth super-chondritc, or the bulk composition of Earth is inherently non-chondritic \citep{ONeill:2008, Caro:2008}. The former explanation has long been popular in accounting for other planetary mass-balance problems \citep{Allegre:1996, Rudnick:2000, Blichert-Toft:1997}, but for the neodymium isotope case it requires an uncomfortably early formation and subsequent isolation of the putative hidden reservoir \citep{Bourdon:2008}. Thus, there is an impetus to explore different models. }

\textcolor{black}{One alternative notes that samarium and neodymium isotopic compositions of different chondrite types show mass independent variations that reflect an inhomogenous distribution of pre-solar materials in the nebular disk \citep{Andreasen:2006}. Some have thus argued that the elevated $^{142}\mathrm{Nd}/^{144}\mathrm{Nd}$ of Earth represents such nucleosynthetic heterogeneity \citep{Huang:2013} rather than the result of radiogenic ingrowth.  However, the few chondrites so far analysed that have Sm and Nd isotopes within the error of measured terrestrial values \citep{Gannoun:2011} have other chemical characteristics that are inappropriate for them to represent the bulk Earth \citep[e.g.][] {Fitoussi:2012}.}

\textcolor{black}{Thus, attention has focused on how collisions during the accretion history of Earth could have altered the Earth's composition from a chondritic starting point \citep[e.g.][]{Palme:2003}. This has many similarities with a hidden reservoir model, except that material is lost from the Earth, rather than irrevocably buried within. If the process of accretion commonly results in bodies with different compositions than the precursor materials, models of planetary compositions require significant revision.}

It is now well documented that planetesimals can differentiate within the first few million years of the Solar System's evolution \citep{Kleine:2005, Schersten:2006, Markowski:2006, Kruijer:2014}. As the terrestrial planets are thought to form from the accretion of planetesimals, it seems inevitable that the planets formed from objects that were already differentiated. Terrestrial planet formation is dominated by collisions, most of which are accretional, but some of which are disruptive. During the course of accretion, compositionally distinct parts of the differentiated colliding bodies might be preferentially lost \citep{Marcus:2009, Marcus:2010}. Indeed, collisional loss of crust during planetary formation has been invoked to be the cause of super-chondritic $^{142}\mathrm{Nd}/^{144}\mathrm{Nd}$ on Earth \citep{ONeill:2008, Caro:2008}. To date, however, attempts to investigate this suggestion quantitatively have been minimal. In part this has been a consequence of the assumption of perfect merging in traditional $N$-body collision models \citep[e.g.][]{Chambers:2001,Kokubo:2002,OBrien:2006,Raymond:2009}. 

\textcolor{black}{In this work we} take a first step towards examining the chemical consequences of imperfect accretion using the parameterisation of collisional out-comes of \citet{Leinhardt:2012}, coupled with the $N$-body gravity code PKDGRAV \citep{Richardson:2000, Stadel:2001}. Rather than address the isotopic differences in $^{142}\mathrm{Nd}/^{144}\mathrm{Nd}$, which rely on rather subtle fractionations between the volumetrically small crust and underlying silicate-dominated mantle, in this \textcolor{black}{initial} study we examine the gross compositional differences between the metallic core and silicate shell (mantle and crust) of planetesimals. Preferential loss of the outer, silicate portion of a planetesimal during collision, increases the fractional mass of its metallic core. Estimates of terrestrial Mg/Fe suggest that it is lower than any chondritic value \citep{Palme:2003}. This has been explained as a consequence of accretional erosion \citep{Palme:2003} but the hypothesis has not been quantitatively tested. Here we examine the importance of preferential mantle erosion during the initial stages of accretion and test if this process can bias the core fraction of planets away from that predicted from the chondritic model.

\section{Numerical Method} \label{sec:num}

\begin{table*}[t]
\caption{Initial conditions for numerical simulations.}
\smallskip
\footnotesize
\begin{tabular}{l | c | l} \hline 
Paramter & Value & Notes\\
\hline \hline
Surface density $\Sigma$ & $\Sigma(r) = 10$ g cm$^{-2} (r/1\,\mathrm{AU})^{-3/2}$ & MMSN\\
\textcolor{black}{Mass of planetesimal disk} & \textcolor{black}{$2.8 M_\oplus$ (resolved) \& $0.028 M_\oplus$ (unresolved)} & from surface density\\
\textcolor{black}{Planetesimal semi-major axis range} & 0.5-1.5 AU & \citet{Kokubo:2002,Leinhardt:2005}\\
Expansion factor ($f$) & $6$ & \citet{Kokubo:2002,Leinhardt:2005}\\
\textcolor{black}{Planetesimal mass density ($\rho$)} & \textcolor{black}{$0.00925$ g cm$^{-3} (f=6)$; 2 g cm$^{-3} (f=1)$} & \citet{Weidenschilling:1977}\\
Eccentricity range & $0 < e < 0.1$ & \citet{Ida:1992, Leinhardt:2005}\\
Inclination range & $0^\circ < I < 10^\circ$ & \citet{Ida:1992, Leinhardt:2005}\\
Velocity dispersion & Rayleigh distribution & \citet{Ida:1992, Leinhardt:2005}\\
\textcolor{black}{Initial $N$} & $10^5$ & \textcolor{black}{$N = 5\times10^5$} at start of perfect merging component\\
Planetesimal radius & 1000-1300 km ($f=6$); 160-215 km ($f=1$) & Perfect merging started with 160 km\\
Timestep & 0.01 yr (major), $1.5\times10^{-4}$ yr (minor) & \citet{Leinhardt:2005}\\
Resolution limit & \textcolor{black}{$2\times10^{22}$ g [1000 km ($f=6$); 160 km ($f=1$)]} & \citet{Leinhardt:2005}\\
Run time & \textcolor{black}{$6 \times 10^5$ yrs ($f=6$); $\sim 2 \times 10^7$ yrs ($f=1$)}  & $6\times10^7$ timesteps\\
\hline \hline
\end{tabular}
\label{tab:init}
\end{table*}

\begin{figure}
\includegraphics[width=0.47\textwidth]{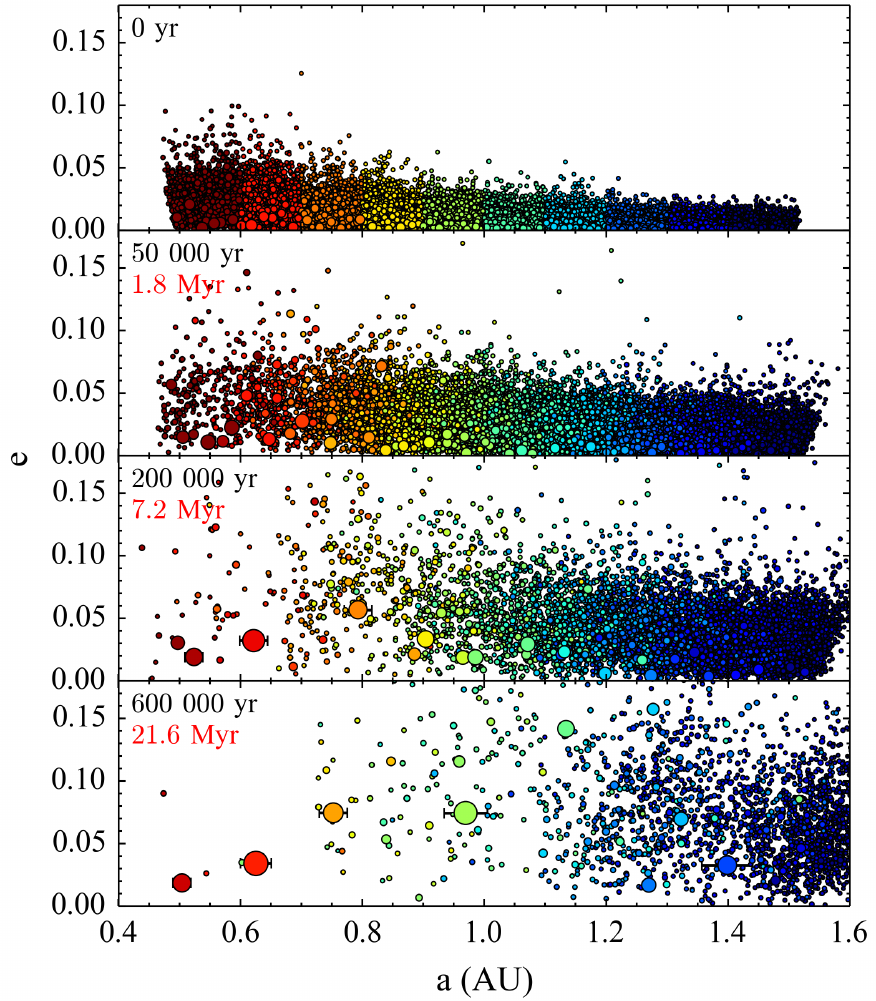}
\caption{\textcolor{black}{Snapshots of the semi-major axis--eccentricity distribution for one simulation, from the initial conditions through to the formation of a chain of protoplanets. The colours indicate the initial location of the material from which the planetesimals are comprised, and can be considered a proxy for composition (as perhaps represented by different chondrite type). The horizontal error bars are 10 Hill radii in length, showing the region of gravitational influence of the protoplanets. The simulated evolution time for the expanded planetesimals ($f = 6$) is indicated in black on each frame the estimated effective evolution time of non-expanded planetesimals ($f=1$) is indicated in red.}}
\label{fig:avse}
\end{figure}

\begin{figure}
\includegraphics[width=0.47\textwidth]{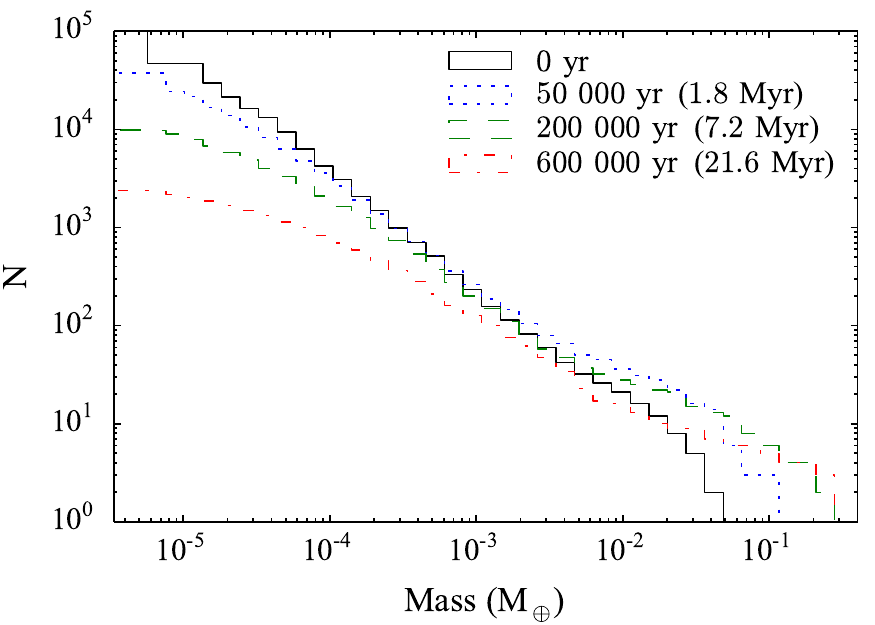}
\caption{The cumulative number of bodies, plotted against their mass, for the same simulation as shown in Fig.~\ref{fig:avse} and plotted for the same four snapshots in time. \textcolor{black}{The effective evolution time is expressed in parentheses.}}
\label{fig:sizedist}
\end{figure}

Terrestrial planet formation is dominated by collisions between planetesimals, both accretional and erosive. In the canonical model of terrestrial planet formation, km-sized bodies grow into protoplanets via runaway and oligarchic growth \citep{Kokubo:1998}. Due to practical numerical constraints, in this work we track the formation of protoplanets from planetesimals that are initially larger than 100 km in diameter.

Our simulations use a modified version of the parallelized $N$-body code, PKDGRAV \citep{Richardson:2000,Stadel:2001}. PKDGRAV \textcolor{black}{employs} a hierarchical tree to calculate \textcolor{black}{inter-particle} gravity and a second-order leap frog integrator for time evolution. The code has been modified to include a state-of-the-art collision model \textcolor{black}{from \citet{Leinhardt:2012}. Unlike many other $N$-body simulations of terrestrial planet formation that assume perfect merging (the projectile is completely accreted by the target) as the only collision outcome \citep[e.g.][]{Kokubo:2002, Raymond:2009}, this code includes a range of collision types in addition to perfect merging, such as, partial accretion (a fraction of the projectile is accreted by the target), hit-and-run (scattering or bouncing collisions in which the target remains intact, although the projectile may be disrupted and no material is exchanged between the target and projectile), and erosive collision outcomes (the target looses mass or is fully disrupted).}

The outcomes of each individual collision, including the size and velocity distribution (in 3D) of the fragments, are calculated based on the impact parameter and collision velocity, using analytic laws derived from simulations of individual collisions \citep{Leinhardt:2012, Stewart:2012}. \textcolor{black}{This collision model produces results that broadly match those of previous work.}

The aim of the simulations in this work is to produce a suite of protoplanets that have the potential to become Earth-like planets. For this reason, and to follow previous work, we focus on the region around 1 AU and use a disc with a surface density of the Minimum Mass Solar Nebula (MMSN) \textcolor{black}{\citep{Kokubo:2002, Leinhardt:2005, Leinhardt:2009}. The mass of the planetesimal disk should not greatly affect our conclusions, having the largest influence on the number of protoplanets formed. }The initial conditions for our simulations are summarised in Table \ref{tab:init}. The simulations are repeated 9 times, with different randomised initial conditions, in order to produce a statistical sample of protoplanets. \textcolor{black}{Each simulation took a few months to complete on} \textcolor{black}{16 2.6 GHz Intel Sandybridge processors}\footnote{https://www.acrc.bris.ac.uk/hpc.htm}.

\textcolor{black}{Due to the vast difference in orbital and collisional dynamical timescales we employ a two rung multistepping procedure. The major step of 0.01 yr is used to resolve the orbit of the planetesimals while a much smaller minor step of $1.5\times10^{-4}$ yr is used to resolve planetesimal collisions \citep[see][for details]{Leinhardt:2005}. However, even with a highly efficient paralleized $N$-body code and multistepping we are forced to ``trick" time in order to reach oligarchic growth in a practical time frame.} Following previous work \citep{Kokubo:2002, Leinhardt:2005, Leinhardt:2009}, we assumed a radial expansion factor\textcolor{black}{, $f$,} of 6 \textcolor{black}{and, thus, an initial planetesimal density ($\rho$) of $0.00925$ g cm$^{-3}$ and a \textcolor{black}{radius} of 1000 km, instead of a density of $\rho = 2$ g cm$^{-3}$ and a \textcolor{black}{radius} of 160 km. As a consequence, the evolution of the planetesimals is accelerated by a factor of $f^2 = 36$. The expansion parameter does not adversely effect the collision model that has been implemented from \citet{Leinhardt:2012} because this model already includes a density normalization (to $\rho = 1$ g cm$^{-3}$); thus, the collision outcome is determined based on the mass of the colliders not the density/radius.}

\textcolor{black}{\citet{Kokubo:1996} showed that radial expansion of this order did not change the growth mode of planetesimals as long as there was still a sufficiently massive background population of smaller planetesimals to dominate the velocity dispersion of all bodies via dynamical friction. In other words, using a radial expansion factor should only change the timescale of evolution as long as the velocity dispersion of the bodies is not dominated by gravitational scattering due to the protoplanets. Thus, our simulations lose validity as we approach the giant impact phase.}

\textcolor{black}{We have chosen to simplify the evolution scenario further, as we have done in the past, by ignoring the presence of any remaining gas disc. This assumption is reasonable for the directly resolved planetesimals as they are large ($\sim100$ km), and the effect of aerodynamic drag is negligible. }

In order to create a \textcolor{black}{more} realistic size distribution than the equal-mass planetesimals used in previous work, and \textcolor{black}{to} avoid an unrealistic number of hit-and-run collisions, the initial size distribution is determined from the outcome of perfectly merging \textcolor{black}{$5\times10^5$ equal-mass planetesimals, until only $10^5$ planetesimals remain}. The initial size distribution is shown \textcolor{black}{as the black line} in Fig.~\ref{fig:sizedist}. 

Our simulations allow erosion, and track the fragments produced in erosive collisions. To stop exponential growth in the number of particles, we do not resolve fragments below a mass of \textcolor{black}{$2\times 10^{22}$ g} ($10^{-11} M_\odot$), which is just under the mass of our initial planetesimals. Fragments smaller than this value are considered to be unresolved material and the mass is added to one of ten cylindrical annuli, depending on the radial location at which the fragments are realised \textcolor{black}{and assumed to be dynamically cold.} The resolved planetesimals continue to accrete this unresolved material as they pass through the annuli. The accretion rate is described by Eq.~4 of \citet{Leinhardt:2005}, and calculated every 10 timesteps, as described in detail in \citet{Leinhardt:2005} and \citet{Leinhardt:2009}. \textcolor{black}{Although the unresolved debris would theoretically be more influenced by a remnant gas disk we have assumed a very simplistic treatment here and again neglected the explicit influence of gas. The influence of gas on the unresolved debris was neglected for several reasons: 1) to stay consistent with previous work; 2) very little unresolved debris was generated at each step and the debris that is generated is placed on a dynamically cold orbit; 3) most of the mass is carried in the largest bodies which have a radius $\sim 100$ km, large enough to be effectively uninfluenced by a remnant gas disk.}

\subsection{\textcolor{black}{Tracking the core mass fraction of protoplanets}}\label{sec:numtrack}

\begin{table*}[t]
\caption{\textcolor{black}{The fate of the core material ($M_{core}$) in the largest remnant of mass $M_{lr}$ from the projectile of mass $M_{proj}$ and core mass $M_{core,proj}$ and target of mass $M_{targ}$ and core mass $M_{core,targ}$ following a collision according to the empirical models (1 \& 2) from \citet{Marcus:2010}. The mantle mass is assumed to make up the rest of the planetesimal mass.}}
\smallskip
\footnotesize
\begin{tabular}{l | l | l} \hline
Collision Type & Model 1 & Model 2 \\
\hline \hline
Perfect Merging & $M_{core} = M_{core,targ} + M_{core,proj}$ & $M_{core} = M_{core,targ} + M_{core,proj}$\\
Partial Accretion & $M_{core} = min(M_{lr}, M_{core,targ} + M_{core,proj})$ & $M_{core} = M_{core,targ} + min(M_{core,proj}, M_{lr} - M_{targ})$\\
\textcolor{black}{Hit-\&-Run (proj.~intact or disrupted)} & $M_{core} = M_{core,targ}$ & $M_{core} = M_{core,targ}$\\
\textcolor{black}{Erosion (partial or supercatastrophic)} & $M_{core} = min(M_{lr}, M_{core,targ} + M_{core,proj})$ & $M_{core} = min(M_{core,targ}, M_{lr})$\\
\hline \hline
\end{tabular}
\label{tab:models}
\end{table*}

The focus of this work is to track the change in the bulk Mg/Fe and Si/Fe ratios between the initially chondritic planetesimals and the protoplanets formed by our $N$-body simulations. These changes occur due to collisions between differentiated bodies \textcolor{black}{ which could cause preferential loss of the outer silicate mantle increasing the proportion of iron-rich core in the object and hence decreasing its Mg/Fe and Si/Fe}. Therefore, we focus initially on the core mass fraction of the planetesimals and protoplanets in our simulations, which is then converted to a bulk composition in \S \ref{sec:collisions}.

In a collision between two differentiated planetesimals, core and mantle material may be spread between the largest remnant and any fragments, depending on the properties of the collision. \citet{Marcus:2009, Marcus:2010} used smoothed particle hydrodynamics (SPH) simulations \textcolor{black}{of} differentiated bodies to follow the outcome of collisions between planetesimals. They tracked the fate of particles that started in the core and mantle of the two colliding bodies, respectively. They find that the core mass fraction of the largest post-collision remnant lies between two extremes; model 1, which over predicts it and model 2, which under-predicts it. The full details of the models are summarised in Table \ref{tab:models}. In model 1 the cores always merge. If the mass of the largest remnant is greater than the sum of the target and projectile cores, mantle is accreted. If the mass of the largest remnant is less than the combined mass of the projectile and target cores, then the remnant will be entirely core. In model 2, cores only merge if the largest remnant is larger than the target. Otherwise in the disruption regime, the mantle is stripped first. \textcolor{black}{In the event there is leftover core material it is distributed to the other collisional fragments starting with the second largest remnant.}

In our simulations, many collisions occur in the history of an individual protoplanet. We post-process each of our simulations six times, using the models of \citet{Marcus:2009, Marcus:2010} in order to determine the change in the core mass fraction of the colliding planetesimals. We assign the same initial core mass fraction to all planetesimals, $c_i$, assuming that they are differentiated at the start of the simulation. We consider three cases, $c_i = 0.35$, representative of high iron content chondrites in a reducing environment, $c_i = 0.1$, representative of low iron chondrites in a \textcolor{black}{more oxidised} environment, and $c_i = 0.22$, a value in the middle of these two extremes. In this manner, once for each of the models of \citet{Marcus:2009, Marcus:2010} and once for each initial core mass fraction, the final core mass fraction is determined for each of the 48 protoplanets formed across our 9 simulations. This gives us a distribution of core mass fractions that could occur for a range of protoplanets formed.

\section{Results of our N-body simulations}\label{sec:results}

\subsection{Comparison with previous work}

The simulations used in this work have the major advantage over many previous simulations \citep[e.g.][]{Kokubo:2002, OBrien:2006, Raymond:2009} that they include many different collision outcomes and follow the evolution of fragments produced in these collisions in a self-consistent manner. The broad results of these simulations, however, match those of previous work \citep[i.e.][]{Kokubo:2002, Leinhardt:2005}. A similar number of protoplanets form over a similar range in semi-major axes, and both the eccentricity distribution (Fig.~\ref{fig:avse}) and mass distribution (Fig.~\ref{fig:sizedist}) of the planetesimals evolve in a similar manner to \citet{Kokubo:2002} or \citet{Leinhardt:2005}. The slope of the mass distribution starts at around -1.5, characteristic of runaway growth. As the simulations progress, the slope decreases, \textcolor{black}{and} the supply of small planetesimals decreases and oligarchic growth dominates, as seen in previous work \citep{Kokubo:2002, Leinhardt:2005}. The major difference between the simulations presented here and previous work is in the timescales. For comparison, in \citet{Leinhardt:2005} oligarchic growth is reached out to 1.5 AU after 400,000 yrs of evolution \textcolor{black}{(in simulated time)}, whereas in our simulations, after 400,000 yrs, oligarchic growth has only reached just beyond 1 AU. Protoplanets form more slowly at larger distance from the star, due to the longer orbital timescales. We include a higher number of particles, $10^5$ compared to $10^4$ in \citet{Leinhardt:2005}, which is equivalent to starting with smaller planetesimals. 

\subsection{The collision history of a protoplanet}\label{sec:collision_history}

\begin{figure}
\includegraphics[width=0.47\textwidth]{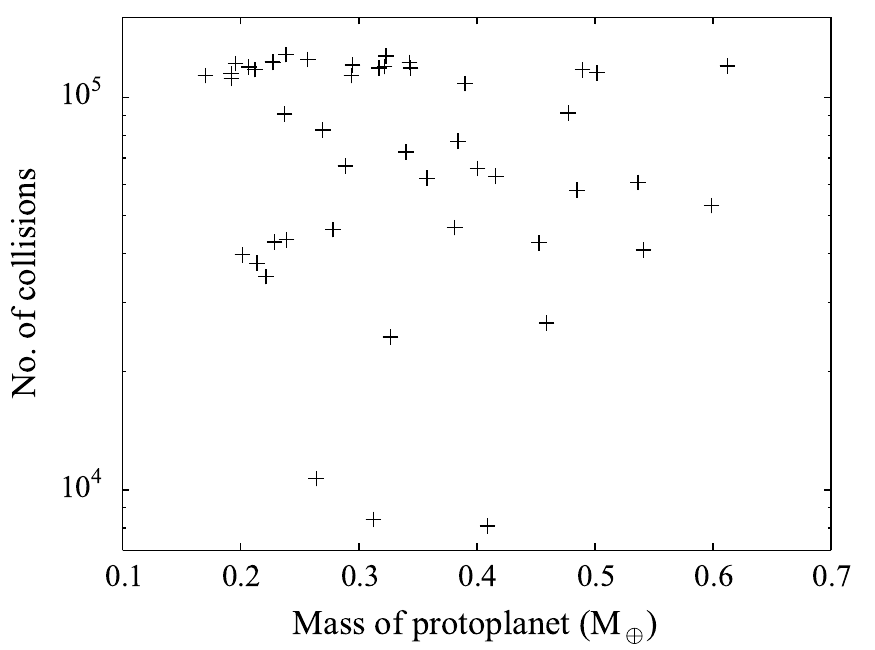}
\caption{\textcolor{black}{Total number of collisions in the history of each protoplanet.}}
\label{fig:coll_proto}
\end{figure}

\begin{figure*}
\includegraphics[width=1.0\textwidth]{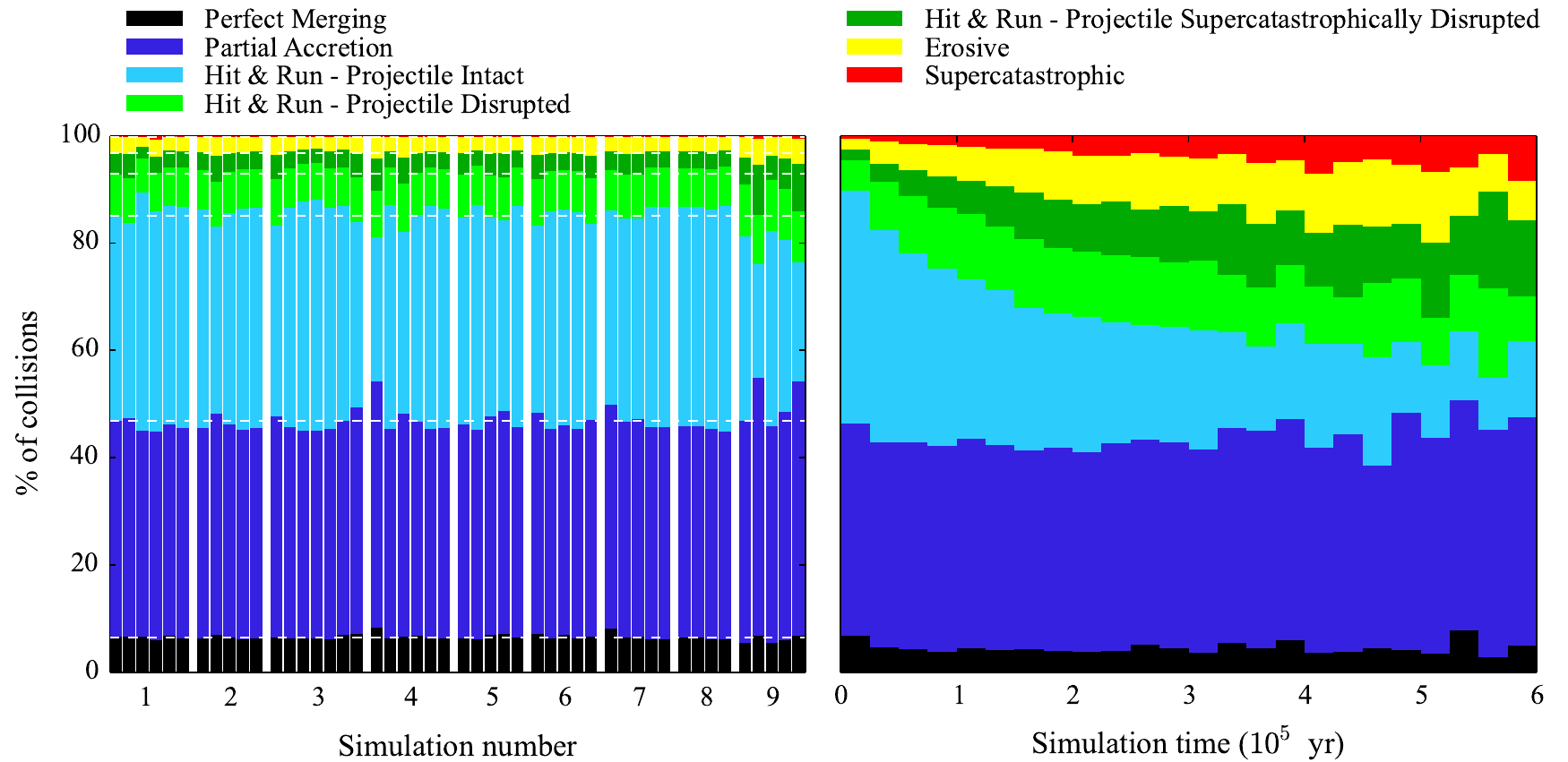}
\caption{\textcolor{black}{The collision history of the simulations, in terms of collision type. The left panel shows the history for each of the 48 protoplanets, each bar representing an individual protoplanet and the white dashed lines the mean. White vertical bars separate individual simulations. The right panel shows the collision history as a function of simulation time (counted in bins of width $2.5 \times 10^4$\,yr) for one simulation. Black indicates perfect merging, blue partial accretion (the target gains mass as a result of the collision), light blue, light green, and green are all hit-and-run collisions meaning that the target stays intact, however, only in light blue events does the projectile stay intact. In light green the projectile looses mass but the target is left intact. In collision outcomes colored green the projectile has been supercatastrophically disrupted and lost more than 90\% of its mass. Yellow indicates erosive collisions in which the target looses mass. Red indicates that the collision was supercatastrophic and more than 90\% of the target mass became unbound as a result of the collision. All simulations evolve in roughly the same manner. }}
\label{fig:coll_history}
\end{figure*}

Each individual protoplanet present at the end of our simulations (after $6\times10^7$ time steps, or $5.7\times10^5$ yrs) formed from collisions between many individual planetesimals. If a protoplanet has on average the isolation mass of $0.16~M_\oplus$, then $3\times10^4$ planetesimals of \textcolor{black}{160 km} are required to form it, and thus, at least a similar number of collisions (assuming all collisions to be perfectly merging). Fig.~\ref{fig:coll_proto} shows that the number of collisions per protoplanet is generally larger than this value, although sometimes this back of the envelope calculation is not valid as significant growth occurred during the perfect merging simulations that determined our initial conditions. In order to determine the effect of collisions on the bulk composition of protoplanets, we track the collision history of each individual protoplanet. Fig.~\ref{fig:coll_history} shows the distribution of different collision types in the collision history of each of our 48 protoplanets, as well as the sum of all collisions that occurred in each of our 9 simulations. On average our simulations are dominated by partial accretion and hit and run events, with a small fraction of collisions being disruptive (erosive or super-catastrophic). However, as shown by Fig.~\ref{fig:coll_history}, the fraction of disruptive collisions increases with time. This is because collision velocities increase with time as the protoplanet excites the planetesimal eccentricities. 

\begin{figure}
\includegraphics[width=0.47\textwidth]{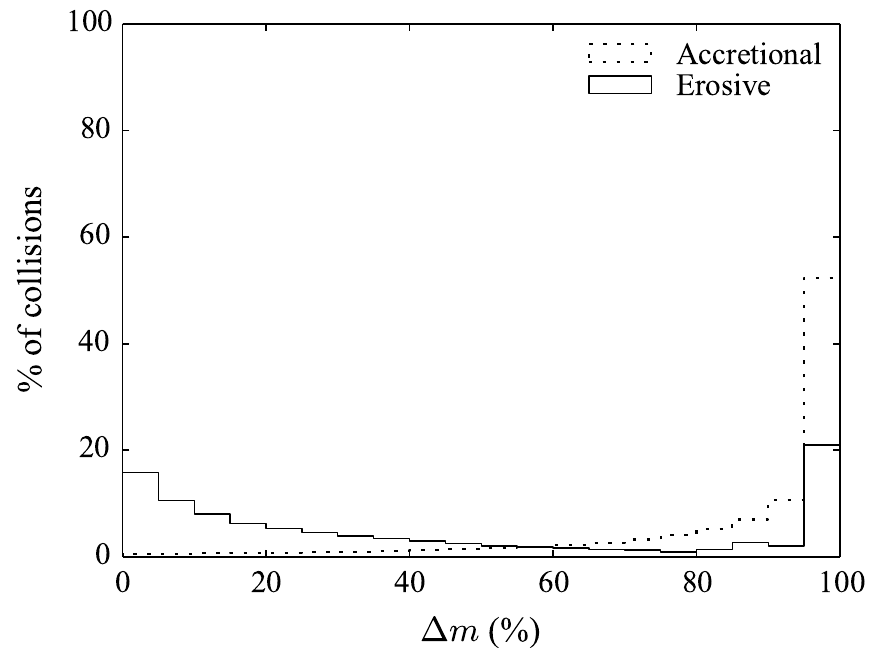}
\caption{The percentage change in mass in all the collisions in our simulations. The solid line shows erosive collisions and $\Delta m_{\mathrm{erosive}}$, whilst the dotted line shows accretional collisions and $\Delta m_{\mathrm{acc}}$, see \S \ref{sec:collision_history} for a full discussion.}
\label{fig:delta_m}
\end{figure}

The majority of collisions in our simulations that are erosive, are not highly erosive. The solid line on Fig.~\ref{fig:delta_m} shows that less than \textcolor{black}{50\% }of the mass is lost in more than \textcolor{black}{half} of all erosive collisions, by considering $\Delta m_{erosive} = \frac{m_{targ}-m_{lr}}{m_{targ}}$, where $m_{targ}$ 
is the mass of the target (the larger of the two colliding bodies) and $m_{lr}$ is the mass of the largest remnant. In a similar manner, most accretional collisions are close to perfect mergers. The dotted line on Fig.~\ref{fig:delta_m} shows $\Delta m_{acc} = \frac{m_{lr}-m_{targ}}{m_{proj}}$, 
where $m_{proj}$ is the mass of the projectile. In over 60\% of collisions, more than 90\% of the total mass available is accreted to form the largest remnant. 

\begin{figure}
\includegraphics[width=0.47\textwidth]{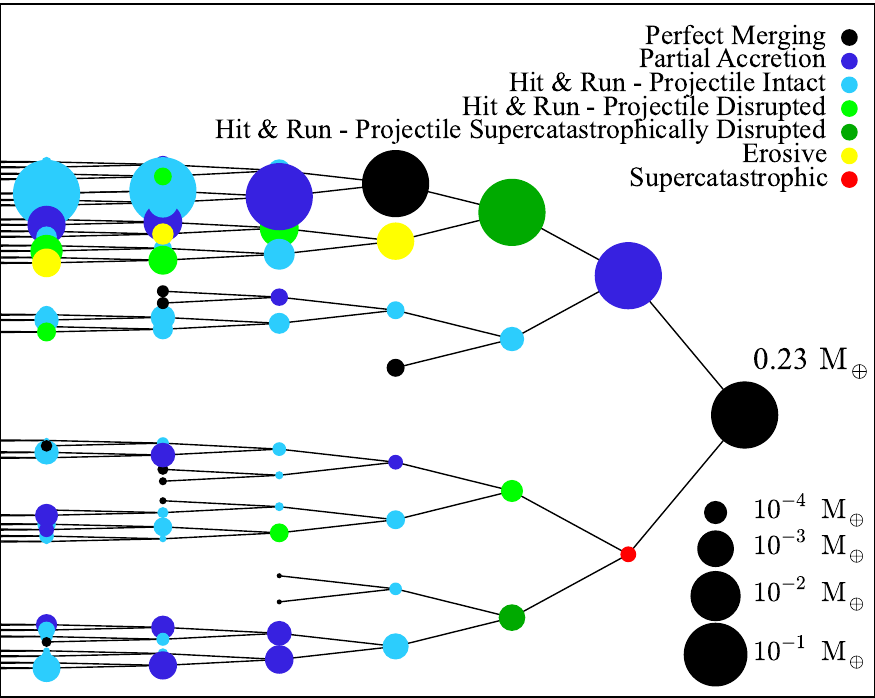}
\caption{An example collision tree \textcolor{black}{of the final collisions involved in constructing a protoplanet.}}
\label{fig:coll_tree}
\end{figure}

A good way to visualise the collision history of an individual protoplanet is via a collision tree. Fig.~\ref{fig:coll_tree} shows an example of some collisions that led to the formation of a \textcolor{black}{$0.23~M_\oplus$} protoplanet. Notice that collisions are presented in order of occurrence and \textcolor{black}{there is no illustration of relative time on this plot}.

\subsection{The change in core mass fraction of a protoplanet}

\begin{figure}
\includegraphics[width=0.47\textwidth]{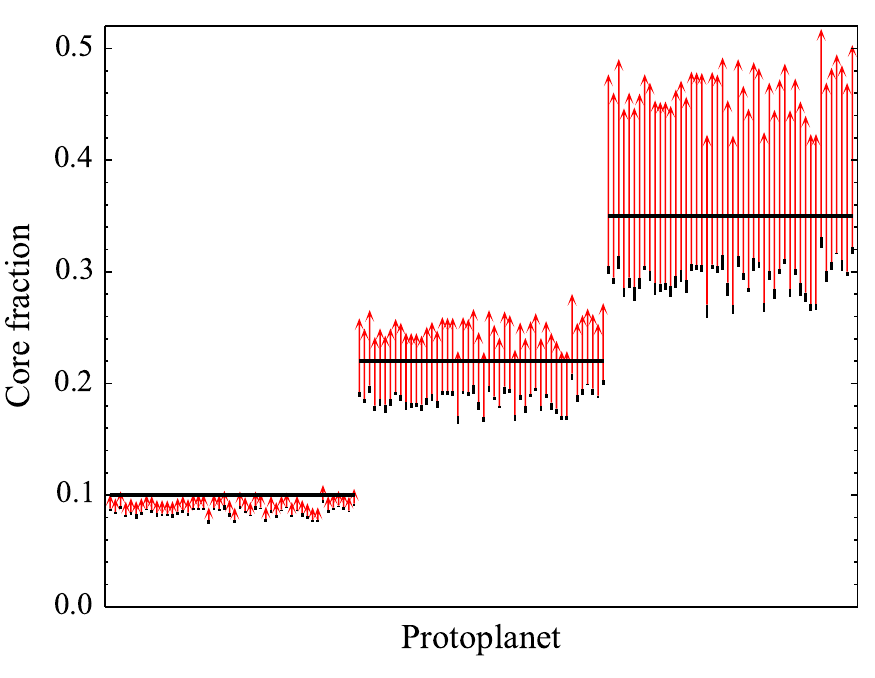}
\caption{\textcolor{black}{The final core mass fraction calculated for each of the 48 protoplanets, formed in 9 simulations. The analysis is
repeated three times, with an assumption of an initial core mass fraction of $c_i$ = 0.1, 0.22 and 0.35. The vertical black bars indicate the variation in core
mass fraction calculated using Model 1 and Model 2, assuming that all unresolved debris is mantle material, whilst the red arrows indicate the
extension to this range, for Model 1, assuming that a fraction of the unresolved debris is core material (see \S\ref{sec:unresolved}).}}
\label{fig:core_mass}
\end{figure}

\begin{figure}
\includegraphics[width=0.47\textwidth]{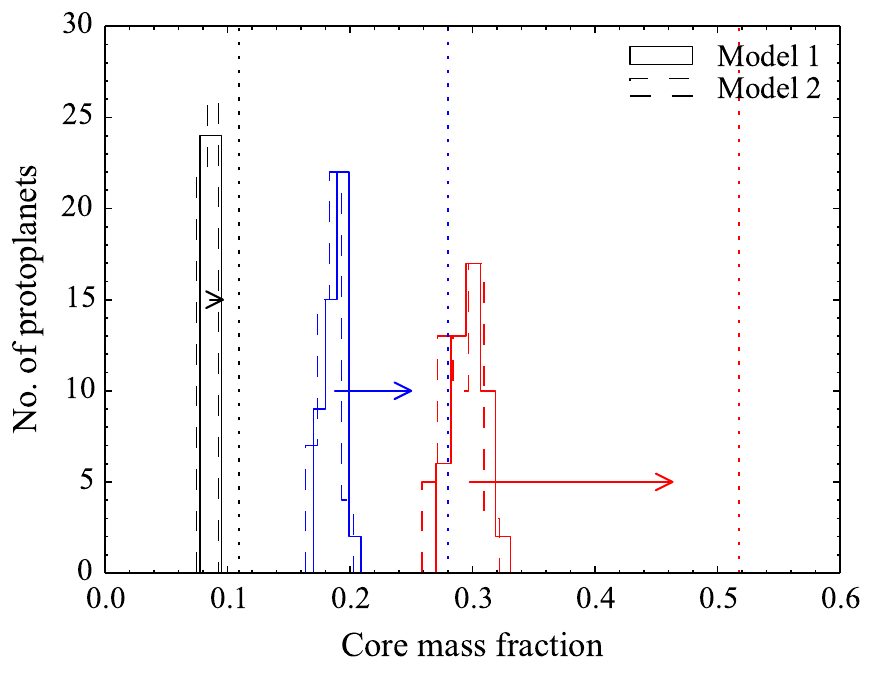}
\caption{A histogram of the change in the core mass fraction found in our simulations, depending on the initial core mass fraction, $c_i$ = 0.1 (black),
$c_i$ = 0.22 (blue) or $c_i$ = 0.35 (red). \textcolor{black}{The solid line indicates Model 1 and the dashed line Model 2. The arrow and dotted line indicate the shift in the mean core fraction, and the maximum core fraction obtained with Model 1 including the error due to the accretion of unresolved material.}}
\label{fig:core_hist}
\end{figure}

As discussed in \S \ref{sec:numtrack}, each of the simulations is post-processed in order to track the final core mass fraction of the protoplanets formed. These are plotted in Fig.~\ref{fig:core_mass}, for all 48 protoplanets formed in 9 separate simulations, analysed three times for initial core mass fractions of $c_i = 0.1$, $c_i = 0.22$ and $c_i = 0.35$ and twice for each model referring to the evolution of the core mass fraction, namely, Model 1 and Model 2. The \textcolor{black}{vertical} black bars indicate the potential range of the final core mass fraction based on the variations between Model 1 and Model 2, assuming that all the unresolved debris is mantle material, whilst the red arrows indicate an extension to this range, assuming that \textcolor{black}{some fraction of this accreted, unresolved debris is core material} (see \ref{sec:unresolved}). This plot clearly shows that in our simulations protoplanets form with a range of different core mass fractions, above and below, the initial core mass fraction of the planetesimals. The core mass fractions referred to here might change as the protoplanets re-equilibriate to their new bulk composition (see discussion in \S \ref{sec:discussion}). Fig.~\ref{fig:core_hist} shows a histogram of these final core mass fractions for all the protoplanets formed in our simulations. The range in final core mass fraction is greater for higher initial core mass fractions. This is because it is easier to disrupt the core of a planetesimal with a larger core mass fraction.

\subsection{Accretion of unresolved material}\label{sec:unresolved}

We limit the resolution of our simulations in order to retain the computing time at a reasonable value. This means that any fragments that are produced in collisions with a mass less than the resolution limit ($10^{-11} M_\odot$) are not followed directly. Instead their mass is added to the mass of unresolved material that is tracked at every timestep. The unresolved material is assumed to form a symmetric, low eccentricity disc (as described in \S \ref{sec:num}) and is accreted by the growing planetesimals every 10 timesteps. This material forms a non-negligible contribution to the total mass of the final protoplanets. In many cases the protoplanet has accreted (as well as eroded) \textcolor{black}{a large fraction of }its final mass in unresolved material during its lifetime. The majority of the unresolved material will be material of mantle composition, but in some cases, particularly where collisions are highly \textcolor{black}{erosive}, large amounts of core material may end up in unresolved debris. 

\textcolor{black}{Our initial calculations assume that all the accreted unresolved material has the composition of the mantle,} \textcolor{black}{and thus do not conserve core material that is lost to the unresolved debris. In fact, during the course of the entire simulation a significant amount of core material is present in the unresolved component, and will be reaccreted by the planetesimals.} \textcolor{black}{We calculated the fraction of core material in the unresolved debris by tracking both the total mass and mass of core material that became unresolved debris during the full simulation. Table~\ref{tab:corefrac} shows the core mass fraction of the unresolved debris at the end of our simulations; as has already been noted, it is easier to excavate core material from planetesimals with higher core mass fractions. We use these values to estimate the additional error on our calculation of the core mass fraction due to the accretion of unresolved material.}

\begin{table*}[t]
\centering
\caption{\textcolor{black}{Core mass fractions of unresolved debris and corresponding errors on core fractions of protoplanets.}\label{tab:corefrac}}
\smallskip
\footnotesize
\begin{tabular}{l r r r} \hline
Initial core fraction of planetesimals & 0.1 & 0.22 & 0.35\\
Final core fraction of unresolved debris & 0.024-0.034 & 0.054-0.081 & 0.089-0.137\\
Protoplanet core fraction increase, $\delta c$ & 14.5\% & 34.2\% & 56.3\%\\
\hline
\end{tabular}
\end{table*}

\textcolor{black}{If a body accretes $m_{dust}$ of material between two adjacent collisions, the core fraction calculated will decrease from $c_1 = \frac{m_{core}}{m_{body}}$, where $m_{core}$ is the mass of the core and $m_{body}$ the mass of the body, to $c_{2,M} = \frac{m_{core}}{m_{body}+m_{dust}}$. This is shown by the black bars on Fig.~\ref{fig:core_mass}. If instead, a fraction, $f_{dust}$, of the accreted material has the composition of the core, then the core mass fraction decreases to $c_{2,C} = \frac{m_{core}+f_{dust} m_{dust}}{m_{body}+m_{dust}}$. Thus, the error introduced in our calculations by this assumption is,}

\begin{equation}
\delta E = \frac{c_{2,C}-c_{2,M}}{c_{2,M}} = \frac{f_{dust} m_{dust}}{m_{core}}.
\end{equation}

\textcolor{black}{Adding errors in quadrature over the typical $10^5$ collisions in the history of a given protoplanet the core fractions of the protoplanets increase by up to,}

\begin{equation}
\textcolor{black}{\delta c = \sqrt{\Sigma^{10^5}_{N=0} \delta E^2}.}
\end{equation}

\textcolor{black}{Taking an example protoplanet, for all collisions the median of $\frac{m_{dust}}{m_{core}} = 0.02$. Thus, using the median, $\delta E = 0.02 f_{dust}$, and taking the lowest final values for $f_{dust}$ found in the 9 simulations, we obtain the values for $\delta c$ listed in table~\ref{tab:corefrac}. In other words, the core mass fractions of our protoplanets could be significantly higher (by $\delta c$) than those we initially calculated, due to the accretion of unresolved material. Our increased estimates of the final core mass fractions are shown by the red arrows in Fig.~\ref{fig:core_mass} and the dotted lines in Fig.~\ref{fig:core_hist}.}

\subsection{Comparison with giant impact simulations}\label{sec:compare}

In this work we investigate whether a change in the core mass fraction, and thus metal-silicate ratio, occurs during the early stages of planet formation, namely, during runaway growth and the onset of oligarchic growth. Such a change could also occur during the final, stochastic stages of planet formation, often referred to as giant impacts. This was previously investigated by \citet{Stewart:2012}, using simulations that start from Mars-mass planetary embryos. \citet{Stewart:2012} use the same collision model as this work, alongside the same prescription for the change in core mass fraction following a collision \citep{Marcus:2009, Marcus:2010}. However, they do not run self-consistent simulations as here, rather, they re-analyse previous simulations from \citet{OBrien:2006} and \citet{Raymond:2009} in which every collision was assumed to result in a perfect merger. This neglects any fragments produced and any change in orbit of the post-collision remnants. 

\textcolor{black}{The} level of increase in the core mass fraction during the early stages of planet formation (this work) is similar to that during giant impacts \textcolor{black}{\citep{Stewart:2012}}, although uncertainties regarding the accretion of unresolved material leave large differences in the spread of final core mass fractions. In other words, there is no information here to suggest that changes to the composition of a planet are more likely to occur during either stage. In fact, given that most planets have evolved through both stages, this highlights that a larger \textcolor{black}{breadth} in the core mass fraction is possible \textcolor{black}{by considering both stages together} than by considering either phase individually. \textcolor{black}{The} effects of collisions through runaway, oligarchic growth and giant impacts have the potential to change further the bulk composition of a planet from the material out of which it formed.

\subsection{Summary}

Our collision model has shown that the collisional evolution of differentiated planetesimals leads naturally to the formation of \textcolor{black}{protoplanets} with a range of core mass fractions spread around the initial core mass fraction of the differentiated planetesimals. \textcolor{black}{For the most extreme initial core mass fraction considered ($c_i = 0.35$), we find a maximum increase to $c_f = 0.52$. In a similar manner planetesimals with an initial core mass fraction of $c_i = 0.22$ can lead to the formation of a protoplanet with a maximum core mass fraction of $c_f = 0.28$. An initial core mass fraction of $c_i = 0.1$ can be increased to a maximum of $c_f = 0.11$. These values take into account the accretion of unresolved material in our simulations, assuming that some fraction of this is core material (see \S \ref{sec:unresolved}).} We note that the core mass fractions calculated here should be considered to be representative of the bulk composition of the protoplanets formed, as iron will redistribute between the core and mantle, depending on the availability \textcolor{black}{of oxygen to change the FeO/Fe ratio of the bulk planet}.

\section{\textcolor{black}{Changing the composition of protoplanets}}\label{sec:collisions}

\begin{figure}
\includegraphics[width=0.47\textwidth]{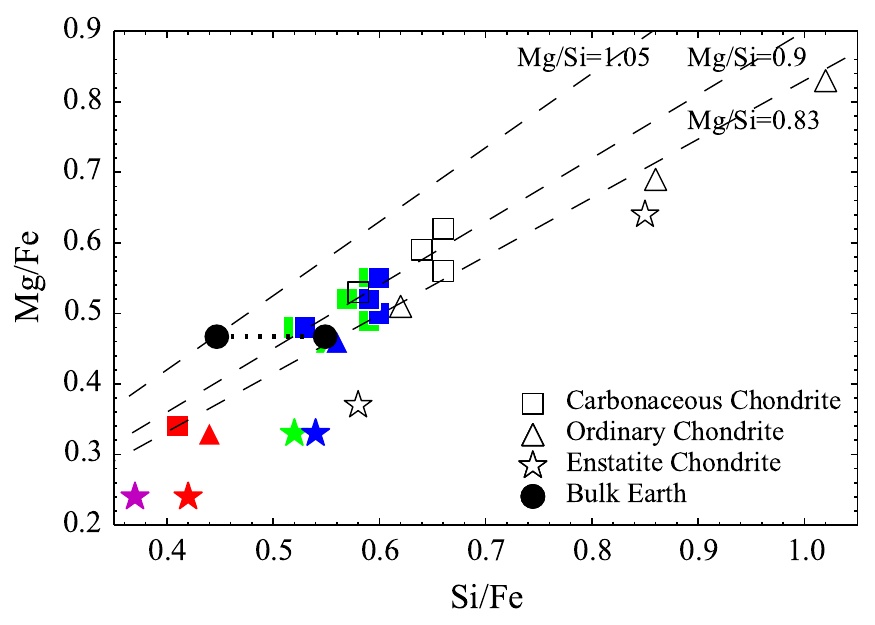}
\caption{\textcolor{black}{The Mg/Fe and Si/Fe ratio for bulk Earth (filled black circles), present day chondrites (open symbols), and chondrites corrected for collisional evolution (filled symbols). The left bulk-Earth data point assumes the core contains light elements (not silicon) whereas the bulk-Earth value on the right assumes that 10\% of the core mass is silicon. The filled shapes show the Mg/Fe and Si/Fe ratios following a collisional history that led to the maximum erosion of the mantle starting from an initial core mass fraction of 0.22 (green) or 0.35 (magenta) assuming no silicon in the core, whilst the blue and red filled symbols show the results from an initial core mass fraction of 0.22 and 0.35, respectively, assuming that the 10\% of the core mass is silicon (see Table \ref{tab:ratios} for tabulated values).}}
\label{fig:earth}
\end{figure}

The Mg/Fe ratio of bulk Earth has been shown to be lower than any chondritic value \citep[e.g.][]{Palme:2003}. Fig.~\ref{fig:earth} shows this ratio for bulk Earth (filled circles) and the chondrites (open shapes), compared to Si/Fe. The composition of bulk Earth can be calculated from the composition of mantle rocks and estimates of the core composition from seismic velocity and density considerations. Here we consider only the \textcolor{black}{six} non-volatile elements (magnesium, aluminium, silicon, calcium, iron and nickel) that dominate Earth's bulk composition \citep{Palme:2003}. Bulk Earth is clearly outside of the range spanned by the chondrites. Bulk Earth must have an excess of iron of at least 10\%, \textcolor{black}{and possibly as much as} 20\%\textcolor{black}{, relative to} the chondrites \citep{Palme:2003}. The collisional erosion of silicates has been suggested as a plausible explanation for this excess \citep{Palme:2003}. In this section we use our analysis of the collisional erosion of differentiated protoplanets in order to assess the viability of this scenario.

\subsection{\textcolor{black}{Tracking Mg/Fe and Si/Fe in the protoplanets}} \label{sec:ratios}

\begin{sidewaystable*}
\centering
\caption{\textcolor{black}{Top: Chondrite compositions in weight percent from \citet{Palme:2003}; Bottom: Protoplanet Mg/Fe, Si/Fe ratios calculated using the method described in \S \ref{sec:ratios} and shown in Fig.~\ref{fig:earth}. `\ldots' indicates that there is insufficient Fe remaining in the protoplanet to match Earth's core mass fraction, `-' indicates that it is not possible to produce a planetesimal with the corresponding initial core fraction. The colors of the text indicate the initial assumed core mass (green and blue indicate $c_i = 0.22$; magenta and red indicate $c_i=0.35$) and the amount of core mass that is assumed to be silicon (green and magenta indicate no silicon in core, blue and red values were calculated assuming 10\% core mass was silicon). The final core fraction ($c_f$) after collisional evolution is given in square brackets.}}\label{tab:ratios}
\smallskip
\footnotesize
\begin{tabular}{c|c|c|c|c|c|c|c|c|c}
\hline \hline
Chondrite Type & CI & CM & CO & CV & H & L & LL & EH & EL\\
\hline \hline
Si & 10.5 & 13.9 & 15.9 & 15.6 & 16.9 & 18.5 & 18.9 & 16.7 & 18.6\\
Al & 0.86 & 1.18 & 1.43 & 1.75 & 1.13 & 1.22 & 1.19 & 0.81 & 1.05\\
Fe & 18.2 & 21 & 24.8 & 23.5 & 27.5 & 21.5 & 18.5 & 29 & 22\\
Mg & 9.7 & 11.7 & 14.5 & 14.5 & 14 & 14.9 & 15.3 & 10.6 & 14.1\\
Ca & 0.92 & 1.27 & 1.58 & 1.9 & 1.25 & 1.31 & 1.3 & 0.85 & 1.01\\
Ni & 1.07 & 1.2 & 1.4 & 1.34 & 1.6 & 1.2 & 1.02 & 1.75 & 1.3\\
O & 46 & 43.2 & 37 & 37 & 35.7 & 37.3 & 40 & 28 & 31\\
\hline
Mg/Fe & 0.53 & 0.56 & 0.59 & 0.62 & 0.51 & 0.69 & 0.83 & 0.37 & 0.64\\
Si/Fe & 0.58 & 0.66 & 0.64 & 0.66 & 0.62 & 0.86 & 1.02 & 0.58 & 0.85\\
\hline \hline
\multicolumn{10}{c}{Results from this work} \\
\multicolumn{10}{c}{ \textcolor{green}{$c_i = 0.22\,[c_f = 0.25]$}, \textcolor{blue}{$c_i= 0.22 (+10\% \mathrm{Si})\,[c_f =0.25]$}, \textcolor{magenta}{$c_i = 0.35\,[c_f=0.46]$}, \textcolor{red}{$c_i = 0.35 (+10\% \mathrm{Si})\,[c_f=0.46]$}} \\
\hline \hline
Initial FeO/Fe & \textcolor{green}{0.57}, \textcolor{blue}{0.64}, -, \textcolor{red}{0.04} & \textcolor{green}{0.46}, \textcolor{blue}{0.52}, -, - & \textcolor{green}{0.45}, \textcolor{blue}{0.52}, -, - & \textcolor{green}{0.39}, \textcolor{blue}{0.45}, -, - & \textcolor{green}{0.55}, \textcolor{blue}{0.63}, -, \textcolor{red}{0.04} & \textcolor{green}{0.23}, \textcolor{blue}{0.29}, -, - & \textcolor{green}{0.075}, \textcolor{blue}{0.13}, -, - & \textcolor{green}{0.75}, \textcolor{blue}{0.83}, \textcolor{magenta}{0.08}, \textcolor{red}{0.16} & \textcolor{green}{0.28}, \textcolor{blue}{0.34}, -, -\\
Final FeO/Fe & \textcolor{green}{0.12}, \textcolor{blue}{0.17}, -, \textcolor{red}{0.54}  & \textcolor{green}{0.05}, \textcolor{blue}{0.09}, -, - & \textcolor{green}{0.05}, \textcolor{blue}{0.09}, -, - & \textcolor{green}{0.01}, \textcolor{blue}{0.06}, -, - & \textcolor{green}{0.11}, \textcolor{blue}{0.16}, -, \textcolor{red}{0.53} & \ldots, \ldots, -, - & \ldots, \ldots, -, - & \textcolor{green}{0.24}, \textcolor{blue}{0.29}, \textcolor{magenta}{0.58}, \textcolor{red}{0.66} & \ldots, \ldots, -, - \\
Final Mg/Fe & \textcolor{green}{0.48}, \textcolor{blue}{0.48}, -, \textcolor{red}{0.34}  & \textcolor{green}{0.49}, \textcolor{blue}{0.50}, -, - & \textcolor{green}{0.52}, \textcolor{blue}{0.52}, -, - & \textcolor{green}{0.55}, \textcolor{blue}{0.55}, -, - & \textcolor{green}{0.46}, \textcolor{blue}{0.46}, -, \textcolor{red}{0.33} & \ldots, \ldots, -, - & \ldots, \ldots, -, - & \textcolor{green}{0.33}, \textcolor{blue}{0.33}, \textcolor{magenta}{0.24}, \textcolor{red}{0.24} & \ldots, \ldots, -, - \\
Final Si/Fe & \textcolor{green}{0.52}, \textcolor{blue}{0.53}, -, \textcolor{red}{0.41}  & \textcolor{green}{0.59}, \textcolor{blue}{0.60}, -, - & \textcolor{green}{0.57}, \textcolor{blue}{0.59}, -, - & \textcolor{green}{0.59}, \textcolor{blue}{0.60}, -, - & \textcolor{green}{0.55}, \textcolor{blue}{0.56}, -, \textcolor{red}{0.44} & \ldots, \ldots, -, - & \ldots, \ldots, -, - & \textcolor{green}{0.52}, \textcolor{blue}{0.54}, \textcolor{magenta}{0.37}, \textcolor{red}{0.42} & \ldots, \ldots, -, - \\
\hline \hline
\end{tabular}
\end{sidewaystable*}

The simulations presented in this work illustrate the manner in which the core mass fraction of a final protoplanet can change significantly from the core mass fraction of the planetesimals out of which that protoplanet formed, as mantle material is preferentially eroded or accreted during collisions. In this section we discuss the implications of this for the bulk composition of the protoplanets formed in our simulations. The aim is to explain Earth's low Mg/Fe ratio. Therefore, we are interested in collisions that preferentially remove mantle material, containing magnesium, over core material, containing most of the \textcolor{black}{proto}planet's iron, as well as planetesimals that start with an initially high iron content. In other words, the focus is on our results for initially high core mass fractions ($c_i = 0.35$, 0.22) and the increase to these core mass fractions seen in our simulations (i.e.~$c_i = 0.35$ can be increased to \textcolor{black}{$c_f = 0.52$ or an average maximum of $0.46$ when corrected for iron in the unresolved debris, and $c_i = 0.22$ can be increased to $c_f = 0.28$ or 0.25 on average}).

\textcolor{black}{Tracking the evolution of the bulk composition of a growing protoplanet is complex and under constrained, thus, we constructed a straight forward but simplified logic. We will first give a brief overview of the 4-stage process and then discuss steps 2-4 in more detail below: 1) Constrain the composition of the mantle and core by assuming that our initial planetesimals all start with the same composition and that this composition is derived entirely from one of the chondrite types (compositions listed in Table \ref{tab:ratios}); 2) Consider the oxidation state of the differentiated planetesimals to be a free parameter allowing us to match the assumption in step 1) with the initial core mass fractions specified in our simulations. Thus, the \textcolor{black}{FeO/Fe ratio} is varied in order to obtain the initial core mass fraction, as shown in Table \ref{tab:ratios}; 3) Calculate the fraction of mass removed to give final core fraction found from the simulations; 4) Change FeO/Fe to attempt to match current Earth core fraction of 32\%.}

The \textcolor{black}{seismologically constrained} density deficity requires the Earth's core to contain $\sim10\%$ light elements \citep[e.g.][]{Birch:1964, Poirier:1994,Allegre:1995,Hillgren:2000}. \textcolor{black}{We consider two extreme possibilities to match this constraint.} In the first, the light element content of the core is entirely elements, such as hydrogen or carbon that are not included in our \textcolor{black}{mass balance} analysis. 
In the second scenario, 
10\% of the mass of the core is assumed to be entirely silicon. This is a reasonable upper limit on the potential silicon content of the core \citep{Hillgren:2000, Ricolleau:2011}. \textcolor{black}{However, the recent work of \citet{Badro:2014} indicate that density and sound velocities calculated for core-forming alloys constrain Si contents to $\sim 5\%$ or less in Earth's core.} An initial core mass fraction of $c_i = 0.35$ can only be obtained from a select sample of chondrite types (CI, H and EH) in the latter case \textcolor{black}{(see Fig.~\ref{fig:earth} and Table \ref{tab:ratios})}.

Using the initial \textcolor{black}{FeO/Fe ratio}, the total iron content of the differentiated planetesimals is divided between FeO in the mantle and metallic iron in the core, $M_{Fe,metal} = M_{Fe,total}/(1 + [\frac{FeO}{Fe}])$, where $M_{Fe, metal}$ is the molar fraction of metallic iron in the core and $M_{Fe, total}$ is the total molar fraction of iron. All silicon, aluminium, magnesium and calcium are assumed to be found in the mantle, except for the case where 10\% of the core mass is assumed to be silicon. \textcolor{black}{All nickel is assumed to be found in the core.} We then consider the potential erosion of the mantle of these differentiated planetesimals. For example, for $c_i = 0.22$, \textcolor{black}{15\% ($\Delta$) of the mantle is eroded, such that the new core mass fraction is given by $c_f = 0.25$.} In general:

\begin{equation} 
\Delta = \frac{M_{mantle,f} - M_{mantle,i}}{M_{mantle,i}} = \frac{c_f - c_i}{c_f (1 - c_i)}
\end{equation}

In order to calculate the bulk composition of the final protoplanet, a fraction, $\Delta$, of the original mass of silicates is removed. If 10\% of the core is assumed to be silicon, then this material is not considered when calculating the removal of material from the mantle. The new ratios of Mg/Fe and Si/Fe for the bulk composition of the protoplanets formed in our simulations are listed in Table \ref{tab:ratios} and shown by the filled shapes on Fig.~\ref{fig:earth}. If the final protoplanets are to have a core mass fraction similar to Earth's 32\%, a new value for the FeO/Fe must be determined. These are listed in Table \ref{tab:ratios} noting that it is not always possible to form a planet with a core mass of 32\%, as indicated by the blanks in Table \ref{tab:ratios}. \textcolor{black}{This indicates that it is more difficult to produce a protoplanet with iron content as high as that of bulk Earth with lower iron content chondrites (such as CV, L, LL and EL) but not impossible if the initial core fraction was high.}

\subsection{Is collisional erosion sufficient to obtain Mg/Fe of bulk Earth from the chondrites?}

The filled shapes on Fig.~\ref{fig:earth} indicate the maximum difference that collisional erosion could make to the Mg/Fe and Si/Fe ratios of a protoplanet formed out of each type of chondrite. The maximum occurs for an initial core mass fraction of $c_i = 0.35$, although a differentiated planetesimal with this core mass fraction can only be formed out of EH, H or CI chondrites, with the assumption that 10\% of the core is composed of silicon. Thus, $c_i = 0.22$ is also considered to be a reasonable maximum and plotted for all chondrite types.

Fig.~\ref{fig:earth} indicates that it is just possible to reduce Mg/Fe to similar values to bulk Earth, particularly if $c_i = 0.35$ and the initial planetesimals start with the composition of the chondrites with the lowest Mg/Fe ratios, i.e. H, EH or CI chondrites. \textcolor{black}{These ratios are obtained more readily if the proto-planet's core contains silicates. Our work shows that} it is possible for collisional erosion during runaway and oligarchic growth to change the composition of the chondrites, in terms of their Mg/Fe and Si/Fe into that of bulk Earth.

\section{Discussion}\label{sec:discussion}

In this work we present $N$-body simulations for the collisional growth of protoplanets, from differentiated planetesimals, in which we constrain the range of core mass fractions expected for a population of protoplanets, based on their initial core mass fraction. If the initial differentiated planetesimals are formed entirely from chondritic material of a given type, we calculate a reasonable maximum for the decrease in the Mg/Fe and Si/Fe ratios, due to the erosion of silicate material from the mantle. We show that a protoplanet with the Mg/Fe and Si/Fe ratios of bulk Earth lies to the extreme of the distribution of protoplanets produced, if Earth's core is assumed to contain 10\% silicon. Without any silicon in the core, Mg/Fe and Si/Fe of bulk Earth are hard to produce. Given the uncertainity in the composition of \textcolor{black}{light elements} in Earth's core \citep[e.g.][]{Hillgren:2000,Ricolleau:2011}, we can reasonably conclude that it is just possible to produce a protoplanet with Earth's Mg/Fe and Si/Fe ratios, from chondritic material, due to collisional erosion of differentiated planetesimals during the early stages of planet formation.

This work, however, is only able to produce a protoplanet of Earth's non-chondritic composition, if Earth happens to be a special case, falling right to one extreme of the distribution of protoplanets that might be formed in this manner\textcolor{black}{, and with a core containing a very high and possibly unrealistic silicon content.} Although Earth may be a special case, there is evidence for a non-chondritic composition among the other terrestrial planets \citep[e.g. Sm/Nd for Mars][]{Caro:2008}. This would suggest that collisional erosion of differentiated planetesimals is commonplace. There are two critical factors that have not been included in the current simulations that may move a protoplanet of Earth's composition more to the centre of the distribution of protoplanets produced. Firstly, we have ignored the presence of gas giants in the early Solar System. Secondly, this work only included collisional evolution during runaway and oligarchic growth, neglecting the giant impact phase at the end of Earth's evolution.

The early evolution of the gas giants in the Solar System may have had a significant influence on the terrestrial planet region. If Jupiter migrated inwards, before being caught in the 2:3 resonance with Saturn and migrating back out to its current position, as in the Grand Tack scenario \citep{Walsh:2011}, it may have caused significant disruption in the terrestrial planet region. Firstly, increasing collision velocities such that the fraction of collisions that are erosive or disruptive increases and more mantle material is likely to be removed. Secondly, scattering material and potentially removing it from the terrestrial planet forming region. Both of these effects have the potential to increase the maximum amount of mantle material that can be stripped from a protoplanet, and thus, potentially making it easier to produce protoplanets with the composition of bulk Earth. This will be investigated in detail in future work.

The collisional evolution described in this work would have continued beyond the formation of protoplanets, albeit in a more stochastic manner, during the giant impact phase. As discussed in \S \ref{sec:compare}, comparison with previous work suggests that the change in composition of the protoplanets during the giant impact phase is similar to that during the earlier stages of growth. We envisage that the cumulative effect of collisional erosion during runaway and oligarchic growth, followed by giant impacts, have the potential to produce planets with a larger range of compositions than those presented in this work. This could mean that planets of Earth's composition are more commonly produced.

We note here one fundamental limitation to the simulations presented here. Although, as discussed in \S \ref{sec:unresolved}, we have made our best attempts to account for any uncertainities, we are limited by the resolution and our inability to follow collisional fragments below a certain size. Our technique of tracking the accretion of unresolved material, of which \textcolor{black}{a few percent} is core material, provides a very good estimate to the behaviour of the system. It does, however, miss any behaviour of these collisional fragments that is specific to the dynamics of a given collision, and the exact composition (mantle/core) of the re-accreted material, may differ from the simple estimates made here.

\section{Conclusions}\label{sec:conclusions}

We investigated whether the non-chondritic composition of Earth, in particular its low Mg/Fe ratio, could be explained due to the collisional stripping of mantle material during the formation of proto-Earth. We focus on the formation of protoplanets from differentiated planetesimals of $>100$ km in size, up to isolation mass. We make use of a state-of-the-art $N$-body code, PKDGRAV \citep{Richardson:2000, Stadel:2001, Leinhardt:2005}, that includes the effects of \textcolor{black}{accretion and erosion in collisions} with a variety of collision outcomes possible (including partial accretion, hit-and-run, supercatastrophic disruption etc). Collisions between differentiated planetesimals are modelled using the results of SPH simulations \citep{Marcus:2009, Marcus:2010}. In this manner, we track the change in the core mass fraction of a protoplanet as it forms. For an initially high initial value of $c_i = 0.35$, we find a spread of final values, with a maximum \textcolor{black}{increase to $c_f = 0.52$}. Or similarly, for an initial value of $c_i = 0.22$, the maximum final value obtained is \textcolor{black}{$c_f = 0.28$}. These calculations take into account the accretion of unresolved material, a necessary feature of our simulations, due to computational limits on the resolution.

If proto-Earth formed from differentiated bodies composed of purely chondritic material, the composition of the mantle and core can be determined by considering the oxygen fugacity of the body to be a free parameter. This model was used to track the maximum change in the bulk composition of a protoplanet, as its core mass fraction changed. Fig.~\ref{fig:earth} shows that the collisional history of a body can decrease the ratio of Mg/Fe sufficiently to retrieve bulk Earth, for planetesimals that started with the composition of the chondrite types with the lowest Mg/Fe (i.e. H, EH or CI chondrites). Si/Fe is only decreased to be \textcolor{black}{comparable} to bulk Earth, if bulk Earth contains 10\% silicon in the core \citep[e.g.][]{Hillgren:2000, Ricolleau:2011}\textcolor{black}{, but which may be unrealistically high \citep{Badro:2014}}. In this case a protoplanet with the composition of bulk Earth lies to the extreme of the distribution of protoplanets produced in our simulations. In other words, Earth would need to be a `special' case, if its non-chondritic composition was derived solely from collisions in the protoplanet's early evolution. We suggest that a combination of collisional evolution prior \textcolor{black}{to} (this work), and during the giant impact phase, both contribute to changes in the composition of Earth from the composition of the disc out of which it formed. Comparison of our work with \citet{Stewart:2012} suggest that both phases can contribute in an equal manner. Inclusion of gas giants into our work may increase the contribution of collisions during the early phases. We are unable to rule out alternative explanations for the non-chondritic composition of Earth, such as a hidden reservoir \citep{Boyet:2005, Labrosse:2007} or heterogeneous proto-solar disc \citep{Andreasen:2006, Huang:2013}. However, in this work, we show that the stripping of mantle material from differentiated planetesimals during the formation of a protoplanet, like Earth, has the potential to change the composition of that protoplanet from the composition of the differentiated planetesimals out of which it formed. This could explain the low Mg/Fe ratio of bulk Earth.

\section{Acknowledgements}

\textcolor{black}{The authors would like to acknowledge useful discussions with D. C. Richardson, A. J. Young, J. Dobinson and S. Lines and helpful reviews from D. O'Brien and an anonymous referee. AB, ZML, PJC, TE, and MW are grateful to NERC grant NE/K004778/1. ZML also acknowledges support from STFC Advanced Fellowship program.}



\bibliographystyle{elsarticle-harv} 
\bibliography{MakingTheEarthBib}


\end{document}